\documentclass[10pt,conference]{IEEEtran}

\usepackage{cite}
\usepackage{amsmath,amssymb,amsfonts}
\usepackage{algorithmic}
\usepackage{textcomp}
\usepackage{xcolor}
\usepackage{graphicx}
\usepackage{inconsolata}
\usepackage{booktabs}
\usepackage{caption}
\usepackage{array}
\usepackage{url}
\usepackage{makecell}
\usepackage{tabularx}
\usepackage{multirow}
\usepackage[table]{xcolor}
\usepackage{threeparttable}
\usepackage[most]{tcolorbox}

\newcommand{\feat}[1]{\texttt{#1}}
\def\BibTeX{{\rm B\kern-.05em{\sc i\kern-.025em b}\kern-.08em
    T\kern-.1667em\lower.7ex\hbox{E}\kern-.125emX}}

\definecolor{ink}{HTML}{2A2E35}
\definecolor{bodybg}{HTML}{FFFFFF}
\definecolor{framec}{HTML}{E5E3DE}
\definecolor{titlebg}{HTML}{F6F4EF}
\definecolor{titleink}{HTML}{33312B}
\definecolor{titlemute}{HTML}{9A958A}
\definecolor{gold}{HTML}{FDE68A}
\definecolor{dotc}{HTML}{FCD34D}
\definecolor{tickc}{HTML}{B0ACA3}

\definecolor{poscol}{RGB}{56,142,60}
\definecolor{negcol}{RGB}{211,47,47}

\makeatletter
\newcommand{\linebreakand}{%
  \end{@IEEEauthorhalign}
  \hfill\mbox{}\par
  \mbox{}\hfill\begin{@IEEEauthorhalign}
}
\makeatother

\makeatletter
\let\old@maketitle\@maketitle
\def\@maketitle{\old@maketitle\vspace{-2em}}
\makeatother

\begin{document}

\title{What Makes a Good Bug Report for an AI Agent?}

\author{
\IEEEauthorblockN{Lara Khatib}
\IEEEauthorblockA{
\textit{University of Waterloo}\\
Waterloo, Canada \\
lara.khatib@uwaterloo.ca}
\and
\IEEEauthorblockN{Noble Saji Mathews}
\IEEEauthorblockA{
\textit{University of Waterloo}\\
Waterloo, Canada \\
noblesaji.mathews@uwaterloo.ca}
\and
\IEEEauthorblockN{Meiyappan Nagappan}
\IEEEauthorblockA{
\textit{University of Waterloo}\\
Waterloo, Canada \\
mei.nagappan@uwaterloo.ca}
\linebreakand
\IEEEauthorblockN{Pengyu Nie}
\IEEEauthorblockA{
\textit{University of Waterloo}\\
Waterloo, Canada \\
pynie@uwaterloo.ca}
\and
\IEEEauthorblockN{Thomas Zimmermann}
\IEEEauthorblockA{
\textit{University of California, Irvine} \\
Irvine, USA \\
tzimmer@uci.edu}
}

\maketitle

\begin{abstract}

Automated program repair (APR) agents are transitioning from research benchmarks to developer workflows, yet they still begin with bug reports written for human developers.
While decades of research have established what makes a good bug report for humans (e.g., steps to reproduce, stack traces), it remains unclear whether these features transfer to LLM-based agents.
We study this question in two complementary analyses.
First, we use statistical modeling to examine associations between 27 bug-report features and repair success across 433 SWE-bench Verified issues attempted by 87 repair agents.
We find that fix suggestions, reproduction scripts, repository source code, and localization info are each associated with higher resolution likelihood, while longer reports are associated with lower odds.
Second, we conduct controlled ablations across 2 models and 17 problem-statement mutations on SWE-bench Pro, systematically varying the information available to an agent while holding the underlying task fixed.
We remove or isolate selected bug-report content and related task information, delete fault-localization cues, and test structural changes that flatten lists or remove section headers without changing the text itself.
By measuring how each change affects agent solve rates, we find that both models depend on localization cues and expected behavior, and that structural changes alone can reduce solve rates, even without removing any content. The two models diverge in how they handle missing information: Qwen searches more widely and can exhaust its turn budget, while Gemma commits to a plausible interpretation early and patches on it.

Our findings indicate that a good bug report for an agent overlaps with, but is not identical to, a good report for a human: 
agents benefit most from concrete, executable, and well-localized information, whereas some qualities long emphasized for human readers, such as natural language steps to reproduce and readable descriptions, contribute little or even correlate with lower success.

\end{abstract}

\section{Introduction}
\label{sec:introduction}

Writing a bug report is often the first step toward resolving a software defect. The report, written in natural language, lays out the problem and guides the investigation that follows. Reporters include information they believe will help identify the cause of the issue, localize the fault, and ultimately produce a fix. A good bug report helps the issue get resolved faster~\cite{zimmermann2009improving}.

LLM-based automated program repair~(APR) agents have moved into real-world engineering workflows, where they read a bug report and submit a patch that attempts to resolve the issue~\cite{yang2024sweagent, deng2025swe}. Yet they still start from the same artifact humans use: the bug report. The agent works through tools that let it read files, run code, and search the repository, with each result becoming part of the context for its next decision. The report determines what information the agent has at the start and shapes the investigation that follows. 

This raises the question: \textbf{\emph{What makes a good bug report for an AI repair agent?}}  
A line of empirical software engineering research has studied what makes a bug report useful for \emph{humans}~\cite{zimmermann2010makes, soltani2020significance, davies2014s}. Prior work has examined which report characteristics developers find most valuable and how report quality relates to triage time, resolution time, and the overall effectiveness of the fixing process~\cite{schroter2010stack, chaparro2019assessing, zimmermann2010makes, soltani2020significance}. This line of work has also shown a mismatch between the information reporters provide developers and the information developers actually find useful when fixing bugs~\cite{zimmermann2010makes, davies2014s}.

However, there is an important difference between an AI agent and a human developer. 
First, what is important for humans fixing a bug may not be as important for AI agents. Existing report templates, reporting guidelines, and bug tracking systems have been optimized for humans~\cite{breu2009frequently, zimmermann2009improving}; however, as AI repair agents take on more of the bug fixing, an equivalent understanding is needed for agents.  
Second, a human developer who receives an incomplete report can ask a follow-up question or collaborate with the reporter~\cite{zimmermann2009improving}. In contrast, current repair agents start from the bug report and run until they either submit a patch or give up, with no guarantee that the patch they produce is correct. Whatever the report fails to provide, the agent must recover on its own or do without.
Therefore, in this paper, we investigate which bug report features contribute to successful resolution by AI repair agents and whether these features align with those previously identified as important for human developers.

Our research consists of two complementary studies.
\begin{itemize}
    \item Study 1 (observational and descriptive): Using the leaderboard from the SWE-bench Verified~\cite{openai2024swebench} benchmark, we collect the outcomes of 87 different agents attempting the same real GitHub issues, annotate each issue for the bug-report features established in prior work, and fit a mixed-effects model that estimates how each feature relates to an agent's probability of resolving the issue, while accounting for the difficulty of the bug and the capability of the agent.
    \item Study 2 (controlled ablation): Starting from solvable issues in SWE-bench Pro~\cite{deng2025swe}, we remove or isolate selected bug-report characteristics and related task information, delete fault-localization cues, and test structural changes that flatten lists or remove section headers without changing the text itself. Because everything else is held constant, each change in solve rate can be attributed to the information removed.
\end{itemize} 

Both studies together help increase validity of this research. Study 1 shows which features are associated with success on real bug reports, which provides ecological validity. Study 2 shows what an agent needs when every other part of the task is held fixed, which provides causal validity; however only under artificial perturbations and not actual real bug reports.

We find that a good bug report for an agent reduces avoidable guessing: it provides executable evidence of the failure, narrows the codebase search, and constrains the required change. These qualities partly overlap with established guidelines for human developers but also diverge: prose steps to reproduce and readable descriptions show no independent benefit, while formatting structure matters independently of content.
In summary, the  main contributions of this paper are:
\begin{itemize}
  \item We investigate what makes a bug report useful to an AI agent, combining statistical analysis of real issue reports with controlled changes and contrasting the empirical findings for AI agents with findings for human developers
  \item A study over 87 agents and 433 real issues that uses a mixed-effects model to estimate the association between each classic bug-report feature and an agent's probability of resolving an issue (Section~\ref{sec:study1_statistical_modeling}).
  \item The first causal experiment to examine which bug-report features affect the performance of AI agents: a controlled ablation across 2 models and 17 problem-statement mutations. We remove or isolate issue components, delete selected information, and remove list formatting or section headers while leaving the remaining text unchanged, and measure how each change affects agent solve rates (Section~\ref{sec:study2_ablation}).
\end{itemize}

\section{Background and Related Work}
\label{sec:background_related_work}

Bug reports communicate defects to developers, and report quality affects their ability to understand and resolve issues \cite{davies2014s}. Prior work consistently finds a gap between what reporters provide and what developers need. Zimmermann et al.~\cite{zimmermann2010makes} identified steps to reproduce, stack traces, and test cases as the most useful elements for fixing bugs. Sasso et al.~\cite{dal2016makes} similarly found that effective elements, including steps to reproduce, defect entities, and test cases, are difficult for reporters to provide, whereas easier elements such as screenshots are less effective. Chaparro et al.~\cite{chaparro2017detecting} confirmed this gap; observed behavior appears in 93.5\% of reports, but expected behavior and steps to reproduce are explicit in only 35.2\% and 51.4\%, respectively. Across interviews with 35 developers, a survey of 305 developers, and 250 GitHub projects, Soltani et al.~\cite{soltani2020significance} likewise found crash descriptions, steps to reproduce, and stack traces important for debugging and associated with bug resolution time. This work has motivated efforts to detect missing information, assess report quality, and improve reports for developers \cite{chaparro2019assessing, bo2024chatbr}, including agentic systems to reproduce bugs and request clarifications \cite{torun2025past}. 

AI repair agents are now entering software development workflows \cite{li2025rise}. They navigate codebases, execute tools, and submit patches autonomously \cite{liu2024large}, but unlike developers who can clarify incomplete reports with reporters \cite{breu2010information}, they must act on the information available. When descriptions omit information, repair agents often fail to seek clarification and produce incorrect code under ambiguity, reducing Pass@1 by 35--52\% \cite{wu2025humanevalcomm}. Despite this limitation, fully autonomous AI repair agents are increasingly deployed in industry \cite{rondon2025evaluating, maddila2026agentic}. They begin from a bug report, then use tools to inspect the repository and produce a patch. Prior work identifies fault localization as a key bottleneck for these systems \cite{zhang2024autocoderover}, and shows that agent--computer interface design affects repair success independently of model capability \cite{yang2024sweagent}. Evaluating these agents requires benchmarks that connect real bug reports to verifiable patches. SWE-bench \cite{jimenez2024swe} curated real GitHub issues to test AI systems in their ability to solve real issues, with SWE-bench Verified \cite{openai2024swebench} providing a human-curated solvable subset and recent benchmarks like SWE-bench Pro \cite{deng2025swe} continue the same established pattern while addressing identified shortcomings. The bug report, however, remains the sole input these systems receive at the start of every task.
To our knowledge, no prior work has systematically examined which bug report features, established as important for human developers, are similarly important for LLM-based agents, nor whether the features that help agents resolve issues differ fundamentally from those that help humans.

\section{Research Questions}
\label{sec:rqs}

Prior work has established which bug report features developers find most useful when resolving defects~\cite{zimmermann2010makes}. However, it remains unclear whether the same features matter when the developer is an AI repair agent. To address this, we ask:

\medskip
\noindent\textbf{RQ1:} \textit{Which bug report features are associated with successful resolution by an AI repair agent?}

\medskip

To answer this question, we analyze outcomes from 87 agents across 433 SWE-bench Verified~\cite{openai2024swebench} instances using a mixed-effects logistic regression model. 

We complement this analysis with a controlled experiment on the well-specified problem statements of SWE-bench Pro~\cite{deng2025swe}, in which we systematically vary the information available to the agent while holding every other aspect of the task fixed. We then measure how these changes affect the agent's solve rate. This leads us to ask:

\medskip
\noindent\textbf{RQ2:} \textit{How do controlled changes to the content and structure of a problem statement affect agent solve rates?}

\section{Study 1: Statistical Modeling}
\label{sec:study1_statistical_modeling}

\subsection{Study Design and Rationale}

Study~1 is observational. We do not run any agents ourselves; we take outcomes already reported for many agents on SWE-bench Verified, annotate each bug report for the features established in prior work, and fit a model relating those features to whether an agent resolved the issue. This tells us which features are \emph{associated} with success, but association is not causation. Study~1 can surface patterns; it cannot tell us whether a feature actually helps the agent.
We run Study~1 on SWE-bench Verified~\cite{openai2024swebench} because it is the only benchmark with publicly reported outcomes from a large, diverse set of agents attempting the same issues, which supplies data for statistical modeling.
The trade-off is that it predates the training cutoffs of newer models on the leaderboard, so we cannot rule out that some agents saw these fixes during training (this limitation is mitigated in Study~2 in Section~\ref{sec:study2_ablation}).

\subsection{Dataset Construction and Filtering}

At the time of our study, 87 different agents had reported results on the SWE-bench Verified leaderboard~\cite{swebench_verified_leaderboard}. For these agents, we collected the released patches and trajectories associated with their leaderboard evaluations. We applied two filters before analysis. First, we retained only bug-fix instances and discarded feature requests and refactoring tasks~\cite{chaparro2017detecting}. The features we annotate (steps to reproduce, observed behavior, stack traces, etc.) are specific to reports that describe a failure, other issue types do not report a failure and typically do not contain these elements. Second, we removed instances with solution leakage, where the problem statement already reveals the fix. Previous research showed that solution leakage is common in SWE-bench Verified and inflates resolution rates~\cite{aleithan2024swe}. We compared each problem statement to its gold patch and marked a problem statement as leaking if it had the patch. Both the issue-type classification and leakage detection were carried out using the same annotation process described in Section~\ref{sec:study1_statistical_modeling:features}. We excluded the flagged instances. After both filters, $433$ instances remained, corresponding to $433$ $\times$ 87 $=$ $37{,}671$ (instance, agent) outcomes, with an overall solve rate of 47.7\%.
Each outcome is binary, resolved or not, as determined by the official evaluation harness. Following SWE-bench evaluation protocol, we define a bug as solved by an agent if the agent successfully submitted a patch that passed all tests, and unsolved if the agent either submitted a patch that failed testing or produced no patch.

\subsection{Feature Extraction and Annotation}
\label{sec:study1_statistical_modeling:features}

The chosen features are based on prior work on bug report quality, in particular the study by Zimmermann et al.~\cite{zimmermann2010makes} of what makes a good bug report. We extract 27 features, listed and defined in Table~\ref{tab:feature-catalog} and organize them into seven categories. These correspond to commonly studied bug report elements~\cite{zimmermann2010makes, soltani2020significance, fan2018chaff, chaparro2017detecting}, and we extend this set with one additional group, fault localization, which is specific to the LLM-based program repair setting. 19 features are binary, indicating whether a given feature is present or absent in a report, and 8 features are continuous.

The first group captures whether a report contains the \emph{core information} required to understand and reproduce a bug. This includes the observed behavior, expected behavior, and steps to reproduce, which prior work consistently identifies as key elements of a bug report~\cite{zimmermann2010makes, chaparro2017detecting}. 
\emph{Diagnostic evidence} refers to runtime artifacts produced during a failure, such as stack traces and error messages.
\emph{Code artifacts} capture whether the report contains code, including illustrative snippets demonstrating the bug, source code drawn from the repository, test cases or references to existing tests, and standalone reproduction scripts that can be executed to trigger the bug. \emph{Environment and guidance} captures contextual and directive information, including software, dependency, runtime, or operating-system versions, in addition to fix suggestions, and external references such as URLs to related issues, discussions or pull requests.

The next two groups capture readability and report structure, focusing on how the report is written rather than what it contains. \emph{Readability} has been widely studied in prior work, and was found to be an important factor in a report's overall quality~\cite{zimmermann2010makes}. Readability measures how easy or hard a text is to read from surface features such as sentence length and word length. We use seven standard readability measures: Flesch Reading Ease, Flesch-Kincaid, the Automated Readability Index (ARI), Coleman-Liau, Gunning Fog, LIX, and SMOG. We also capture structural signals that influence how information is organized. \emph{Structure} features record whether a report uses lists (itemization) and whether it contains section headers. \emph{Itemization} has been identified in prior work as an important signal of bug report quality~\cite{zimmermann2010makes}. \emph{Section headers} capture whether the report is organized into explicit sections such as “Expected behavior,” “Steps to reproduce,” or similar standard bug report headings. These signals reflect formatting decisions that may affect how easily a report can be scanned and interpreted, even when the underlying content remains the same.  We also include \emph{report length}, which prior work treats both as a proxy for information content and as a potential source of noise when reports become overly verbose.

The final group, \emph{fault localization}, is specific to this study and the agent setting. Fixing a bug requires identifying the relevant location in a large codebase, and prior work on automated program repair has shown that this step is often a key bottleneck in the repair process~\cite{jimenez2024swe, zhang2024autocoderover}. A report that explicitly identifies code modified by the eventual fix may substantially reduce this search effort. We therefore measure whether a report explicitly names code elements modified by the ground-truth patch, including files, modules, classes, or functions. Prior work only notes this signal indirectly, for example through stack traces or patch references~\cite{zimmermann2010makes}.

To construct the feature set, we first generate initial labels automatically. For the content features, we prompt an LLM (GPT-5-mini) to read each problem statement and produce a present-or-absent label for every feature with supporting text from the report. For the fault localization features, we extract the files, modules, classes, and functions changed in the ground-truth patch and match them against the report. 
Two authors then independently reviewed and verified each label using a shared annotation guide. We report inter-annotator agreement with Cohen's $\kappa$, which ranged from 0.92 to 1.00, and resolved disagreements by discussion. The per-feature values appear in the replication package. 
For continuous features, we compute length as the number of characters in the problem statement, and readability using the \texttt{readability} Python package~\cite{readability_python}. Readability is computed after preprocessing the problem statement to remove non-natural language content such as code blocks, stack traces, and patches.

\begin{table*}[t]
\centering
\footnotesize
\caption{Bug report features and metrics used in Study~1. Prevalence is the percentage of reports containing each binary feature ($n = 433$).}
\label{tab:feature-catalog}
\renewcommand{\arraystretch}{1}
\begin{tabular*}{\textwidth}{@{\extracolsep{\fill}} l l p{7cm} r @{}}
\toprule
\textbf{Category} & \textbf{Feature} & \textbf{Description} & \textbf{Prevalence (\%)} \\
\midrule
\textit{Core information}
  & \feat{has\_observed\_behavior}   & Observed, incorrect behavior (OB)                              & 98.8 \\
  & \feat{has\_expected\_behavior}   & Expected, correct behavior (EB)                                & 91.9 \\
  & \feat{has\_steps\_to\_reproduce} & Steps and conditions to reproduce the failure (S2R)            & 58.9 \\
\midrule
\textit{Diagnostic evidence}
  & \feat{has\_stack\_trace}         & Stack trace with file paths and line numbers                  & 20.1 \\
  & \feat{has\_error\_message}       & Verbatim error or exception message                           & 38.8 \\
\midrule
\textit{Code artifacts}
  & \feat{has\_code\_snippet}        & Illustrative snippet demonstrating the bug                    & 81.3 \\
  & \feat{has\_repository\_code}     & Project source code drawn from the repository                 & 12.2 \\
  & \feat{has\_test\_case}           & Executable test code with assertions                          & 5.5 \\
  & \feat{has\_test\_reference}      & Reference to a specific test file or test method              & 3.9 \\
  & \feat{has\_reproduction\_script} & Standalone script that reproduces the failure                 & 48.3 \\
\midrule
\textit{Environment and guidance}
  & \feat{has\_environment\_info}    & Software, dependency, runtime, or OS versions                 & 49.7 \\
  & \feat{has\_fix\_suggestion}      & Candidate fix, change, or workaround                          & 66.7 \\
  & \feat{has\_external\_reference}  & External link to a related issue, discussion, or PR           & 48.5 \\
\midrule
\textit{Readability}
  & \makecell[l]{\feat{flesch}, \feat{kincaid}, \feat{ari}, \feat{coleman\_liau}\\
                 \feat{fog}, \feat{lix}, \feat{smog}}
  & \makecell[l]{Standard readability metrics measuring text complexity\\
                  computed after stripping code from report}
  & -- \\
\midrule
\textit{Structure}
  & \feat{has\_itemization}          & Uses bulleted or numbered lists                               & 10.6 \\
  & \feat{has\_section\_headers}     & Organizes content under explicit section headings             & 28.6 \\
  & \feat{length\_chars}             & Length of the problem statement in characters                 & -- \\
\midrule
\textit{Fault localization}
  & \feat{localizes\_file}           & Names a file modified in the ground-truth patch               & 28.6 \\
  & \feat{localizes\_module}         & Names a module or import path mapping to a modified file       & 14.5 \\
  & \feat{localizes\_class}          & Names a class modified in the ground-truth patch              & 14.1 \\
  & \feat{localizes\_function}       & Names a method modified in the ground-truth patch  & 39.5 \\
\bottomrule
\end{tabular*}
\end{table*}

\subsection{Mixed-Effects Regression Model}

To examine the association between bug report features and agent solvability outcomes, we employed a logistic mixed-effects regression model~\cite{baayen2008mixed}. Mixed-effects regression was selected to account for the crossed structure of the data, in which each instance is attempted by many agents and each agent attempts many instances. We define an \textit{agent} as a unique combination of a scaffolding system and the underlying LLM it uses. The 87 agents in our dataset span a wide range of proprietary and open-weight LLMs with solve rates ranging from 10.2\% to 75.2\%. We include crossed random intercepts for instance and for agent to account for these repeated attempts, the instance intercept gives each bug its own baseline solve rate, and the agent intercept gives each agent its own baseline skill. A random intercept gives each bug its own baseline, but on its own it cannot separate the effect of a report feature from the difficulty of the bug, because harder bugs tend to come with differently written reports. To estimate feature effects while holding difficulty constant, we therefore enter bug difficulty as a fixed-effect predictor, taken from the four-level time-to-resolve rating released with SWE-bench Verified~\cite{openai2024swebench}. Formally:

\begin{equation}
\text{logit}(\Pr(\text{solved} = 1)) =
\beta_0 + \sum_j \beta_j x_j + u_{\text{instance}} + u_{\text{agent}}
\end{equation}

\noindent where $x_j$ are the fixed-effect predictors with coefficients $\beta_j$. The terms $u_{\text{instance}}$ and $u_{\text{agent}}$ are random intercepts for instance and agent, respectively. The fixed effects therefore capture the contribution of bug report features to the probability of resolution after accounting for variation in instance difficulty and agent capability. We fit the model using the \texttt{glmer} function from the \texttt{lme4} package in R~\cite{bates2015fitting}. For each predictor, we report the effect size as an odds ratio (OR), computed as the exponentiated regression coefficient, together with a 95\% confidence interval and a significance indicator. An OR greater than 1 implies that the presence of a feature increases the likelihood of resolution; an OR below 1 indicates the opposite. We additionally report the marginal and conditional $R^2$ of Nakagawa and Schielzeth~\cite{nakagawa2013general}, which quantify the variance explained by the fixed effects alone and by the full model including random effects.

\subsection{Feature Selection and Preparation}

Prior to fitting the model, we examined the instance-level prevalence of each binary feature across the 433 retained instances (Table~\ref{tab:feature-catalog}). Features with prevalence outside the $[10\%, 90\%]$ range were excluded, as near-constant predictors provide insufficient variation for reliable estimation. On this basis, \feat{has\_observed\_behavior} (98.8\%) and \feat{has\_expected\_behavior} (91.9\%) were excluded for near-universal prevalence, and \feat{has\_test\_case} (5.5\%) and \feat{has\_test\_reference} (3.9\%) for insufficient variance. We also fit a model that retains all features, which yielded consistent estimates for the features shared with the primary model, and we report it in the replication package. We included report length and the seven readability metrics as continuous variables. Report length, the number of characters in the problem statement, was strongly right-skewed (median 1,179 characters, maximum 24,770, skewness 6.35), so we used $\log(\text{length})$ in place of the raw count. To identify potential collinearity, we examined correlations among predictors. We evaluated multicollinearity using the Variance Inflation Factor (VIF), excluding predictors with VIF $\geq 5$~\cite{fox1992generalized}. The seven readability metrics measure the same surface features (sentence length and word length), so they were highly correlated and produced VIF values well above this cutoff when entered together. We therefore retained a single readability metric, \feat{smog}, in the primary model and refit with each of the other six in its place. After this reduction, every predictor in the final model had a VIF below 5. We z-standardized \feat{log\_length} and the seven readability metrics. Their odds ratios therefore give the change in odds for a one-standard-deviation increase. Binary features remained coded as zero or one. A binary feature's odds ratio gives the change for that feature being present versus absent. The final model contained 15 binary features, \feat{log\_length}, and \feat{smog}, the 17 features in Table~\ref{tab:glmm-difficulty}.

\subsection{Results}

\noindent\textbf{RQ1: Which bug report features are associated with successful resolution by an AI repair agent?}

\begin{table}[t]
\centering
\caption{Mixed-effects logistic regression results, grouped by category.}
\label{tab:glmm-difficulty}
\small
\setlength{\tabcolsep}{4pt}
\renewcommand{\arraystretch}{1.10}
\begin{tabular*}{\columnwidth}{@{\extracolsep{\fill}}l>{\raggedleft\arraybackslash}p{13mm}ll@{}}
\toprule
\textbf{Feature} & \textbf{OR} & \textbf{95\% CI} & \textbf{Sig} \\
\midrule
\multicolumn{4}{@{}l}{\textit{Core information}} \\
\feat{has\_steps\_to\_reproduce} & 0.83 & [0.47, 1.47] & \\
\addlinespace
\multicolumn{4}{@{}l}{\textit{Diagnostic evidence}} \\
\feat{has\_stack\_trace} & 0.93 & [0.38, 2.29] & \\
\feat{has\_error\_message} & 1.12 & [0.59, 2.16] & \\
\addlinespace
\multicolumn{4}{@{}l}{\textit{Code artifacts}} \\
\feat{has\_code\_snippet} & 1.45 & [0.67, 3.14] & \\
\feat{has\_repository\_code} & \cellcolor{poscol!36}2.82 & [1.23, 6.44] & * \\
\feat{has\_reproduction\_script} & \cellcolor{poscol!32}2.52 & [1.41, 4.51] & ** \\
\addlinespace
\multicolumn{4}{@{}l}{\textit{Environment and guidance}} \\
\feat{has\_environment\_info} & \cellcolor{poscol!19}1.72 & [0.94, 3.17] & . \\
\feat{has\_fix\_suggestion} & \cellcolor{poscol!45}3.61 & [2.01, 6.47] & *** \\
\feat{has\_external\_reference} & 0.72 & [0.43, 1.20] & \\
\addlinespace
\multicolumn{4}{@{}l}{\textit{Readability}} \\
\feat{smog} (z) & \cellcolor{poscol!15}1.55 & [1.18, 2.04] & ** \\
\addlinespace
\multicolumn{4}{@{}l}{\textit{Structure}} \\
\feat{has\_itemization} & 0.82 & [0.34, 1.96] & \\
\feat{has\_section\_headers} & 1.78 & [0.88, 3.62] & \\
\feat{log\_length} (z) & \cellcolor{negcol!25}0.49 & [0.35, 0.68] & *** \\
\addlinespace
\multicolumn{4}{@{}l}{\textit{Fault localization}} \\
\feat{localizes\_file} & \cellcolor{poscol!30}2.33 & [1.18, 4.60] & * \\
\feat{localizes\_module} & 1.04 & [0.52, 2.07] & \\
\feat{localizes\_class} & 1.10 & [0.54, 2.23] & \\
\feat{localizes\_function} & 1.40 & [0.80, 2.44] & \\
\bottomrule
\end{tabular*}
\\[3pt]
{\footnotesize
 \textsuperscript{***}$p<0.001$, \textsuperscript{**}$p<0.01$, \textsuperscript{*}$p<0.05$, \textsuperscript{.}$p<0.10$\par
\smallskip
$\sigma^2_{\text{inst}}=5.98$, \quad $\sigma^2_{\text{agent}}=1.99$, \quad $R^2_{m}=0.24$, \quad $R^2_{c}=0.78$\par}
\end{table}

Table~\ref{tab:glmm-difficulty} presents the results of the mixed-effects logistic regression model. The instance random-effect variance (5.98) is 3$\times$ the agent variance (1.99). This means differences between bugs account for more of the variation in success than differences between agents do. Some bugs are consistently easier across agents and others consistently harder. The marginal $R^2$ is 0.24 and the conditional $R^2$ is 0.78. This means the fixed-effects alone explain about 24\% of the variation in success, and that rises to 78\% once the model accounts for which bug and which agent each attempt involves. The features associated with a higher chance of resolution fall in three of the categories in Table~\ref{tab:feature-catalog}: code artifacts, environment and guidance, and fault localization. A suggested fix (\feat{has\_fix\_suggestion}) had the largest effect, multiplying the odds of resolution by 3.61,  repository source code (\feat{has\_repository\_code}) by 2.82, a reproduction script the agent can run (\feat{has\_reproduction\_script})  by  2.52, and naming a file the fix changes (\feat{localizes\_file}) by 2.33. This means the features that help are the
ones that give the agent a more concrete starting point. Some provide material it would otherwise have to find or write, and others narrow the part of the repository it needs to inspect. Report length (\feat{log\_length}) showed a negative association (OR = 0.49). Holding the other included features constant, a one-standard-deviation increase in log-transformed report length was associated with 51\% lower odds of resolution. Reports with higher \feat{smog} scores, indicating harder prose, had slightly higher odds of resolution (OR~1.55). Refitting with the other six readability measures gave the same direction for all seven, with six significant (results in the replication package). Because readability excludes code, stack traces, and other non-prose content, this pattern reflects the report's natural-language prose. 

Several features commonly emphasized in guidance for human-written bug reports did not show an effect here. These included stack traces, error messages and structure such as itemization. This does not show that these features are unhelpful. Rather, in this dataset, their presence was not associated with a detectable difference in resolution after accounting for the other report features, issue-level differences, and agent-level differences. The two core elements of a bug report, observed behavior and expected behavior, also could not be evaluated in the model. Nearly every retained report stated both what the program does wrong and what it should do instead, leaving too few reports without these elements to estimate their associations reliably. These limits of the observational analysis motivate Study~2 in Section ~\ref{sec:study2_ablation}, which tests the effect of report content under controlled changes to the problem statement.

\section{Study 2: Controlled Ablation}
\label{sec:study2_ablation}

\subsection{Study Design and Rationale}

Study~1 shows which features are associated with success, but not whether removing one would actually hurt an agent. Study~2 answers that with a controlled ablation: we start from real bug reports, delete one category of information at a time, rerun a repair agent on the modified report, and measure the drop in solve rate. Everything else about the task is held fixed, so a drop can be attributed to the content we removed. 

\subsection{Dataset Selection and Filtering}
\label{sec:study2_dataset}

We use the public set of SWE-bench Pro~\cite{deng2025swe}, a benchmark of 731 instances drawn from 11 actively maintained repositories, spanning Python, Go, JavaScript, and TypeScript. 
We do not use SWE-bench Verified as in Study~1 because some of its issues and fixes may be contaminated by the training data of the models we use in Study~2, which could bias the ablation study results~\cite{liang2025swebenchillusion}.
In contrast, SWE-bench Pro is a contamination-resistant benchmark: all repositories are distributed under strong copyleft licenses (GPL), which makes their inclusion in commercial training corpora legally and practically unlikely~\cite{deng2025swe}.
SWE-bench Pro labels each instance with one or more issue types: bug fix, feature request, or enhancement. Applying the same filtering criterion as Study~1, we retained only instances that had a bug-fix label, discarding any instance that carried only a feature-request or enhancement label. Of the 731 public instances, 283 are labeled as bug fixes, spanning all four languages: 112 Go, 107 Python, 51 JavaScript, and 13 TypeScript.

\subsection{Anatomy of an SWE-bench Pro Instance}
\label{sec:anatomy}

Each instance in SWE-bench Pro can be decomposed into five components, where the report content can be further decomposed into four sections, as listed below (the components/sections marked with * are required, others are optional):

\begin{itemize}
    \item \textbf{Title*:} the issue title.

    \item \textbf{Description*:} a summary of the issue.

    \item \textbf{Report Content*:} the body of the issue report that describes the following sections:
    \begin{itemize}
        \item \textbf{Observed Behavior*:} what the program actually does; the incorrect output, error, or symptom.
        \item \textbf{Expected Behavior*:} what the program should do.
        \item \textbf{Steps to Reproduce:} the explicit steps or conditions that trigger the failure.
        \item \textbf{Additional Context:} supporting information such as suggested fix or environment details.
    \end{itemize}

    \item \textbf{Requirements*:} (added by SWE-bench Pro) specifies the behavior that a correct solution must satisfy.

    \item \textbf{Interface:} (added by SWE-bench Pro) specifies the relevant files, classes, methods, or signatures that a correct solution is expected to expose.
\end{itemize}

The title, description, and report content are extracted from the original GitHub issue, representing the information available in a bug report at the time of reporting.
The requirements and interface are augmented by SWE-bench Pro authors, with the goal of capturing the communication among users and developers and guiding the agent to produce the correct solution expected by the benchmark tests~\cite{deng2025swe}.

\subsection{Ablation Design}
\label{sec:ablation_design}

We define four types of mutations. Spec-composition mutations remove one component from the instance or retain that component alone. The localization mutation removes references to files changed by the fix from the instance. Content mutations retain only the report content, then remove sections from it. Structural mutations change the formatting of the instance without removing any text.

\subsubsection{Spec-composition mutations}
The five top-level components are the title, description, report content, requirements, and interface. For each component \textit{X}, we run two conditions:

\begin{itemize}
\item \textbf{All\,$-$\,\textit{X}}: remove component \textit{X} and keep the other four.
\item \textbf{Only~\textit{X}}: keep component \textit{X} and remove the other four.
\end{itemize}

\noindent This gives ten conditions: All $-$ Title, All $-$ Description, All $-$ Report Content, All $-$ Requirements, All $-$ Interface, Only Title, Only Description, Only Report Content, Only Requirements, and Only Interface. Only~\textit{X} shows how often the agent solves the bug from
component \textit{X} alone. All\,$-$\,\textit{X} shows how much the solve rate drops when component \textit{X} is
removed from the full statement.

\subsubsection{Localization mutation}

Localization information is not confined to a single component; a file modified in the gold patch can be named in the report content, the requirements, or the interface. The Remove All File References mutation removes every such reference across the full problem statement. This mutation corresponds to the \texttt{localizes\_file} feature from Study~1 (Section~\ref{sec:study1_statistical_modeling}).

\subsubsection{Content mutations}
Starting from the report content (RC) alone, we remove one section at a time: RC $-$ Observed Behavior, RC $-$ Expected Behavior, RC $-$ Steps to Reproduce, and RC $-$ Additional Context.

\subsubsection{Structural mutations}
Structural mutations change the formatting of the full problem statement and leave the text unchanged. Because the text is identical to the baseline, any change in solve rate comes from the formatting. We run two:

\begin{itemize}
\item \textbf{Remove Section Headers:} remove every section header so the sections run together.
\item \textbf{Flatten Lists:} flatten bulleted and numbered lists into prose,
removing list markers, numbering, and line breaks.
\end{itemize}

\subsection{Experimental Setup}
\label{sec:study2_setup}

We run all experiments with mini-SWE-agent~\cite{miniswagent}, a minimal harness in which the model interacts with the repository through a bash command loop, without a dedicated tool-calling interface and with a linear message history. We chose this harness for two reasons. First, in early testing with SWE-agent~\cite{yang2024sweagent}, tool-layer failures were frequent and instances were recorded as unsolved for reasons unrelated to the problem statement. A simpler harness reduces the number of ways a run can fail for reasons unrelated to the bug report, which is precisely what we want when isolating the effect of report content. 
Second, on the SWE-bench Pro leaderboard at the time of evaluation~\cite{swebench_pro_leaderboard}, mini-SWE-agent is consistently used for newer models (e.g., GPT-5.4) and outperforms SWE-agent with older models (e.g., GPT-5).
We evaluate two models from two families of open-weight models, Qwen3.6-35B-A3B~\cite{yang2025qwen3} and Gemma-4-31B-IT~\cite{gemma42026}. Each run executes in an isolated Docker container with all dependencies installed, under a budget of 50 agent turns and a \$2 cost cap. Every run in this study is repeated three times. LLMs are non-deterministic: the same model given the same prompt does not always produce the same patch and may solve an instance on one run and fail on it the next. With a single run we could not tell whether a failure came from a mutation or from run-to-run variation.

\subsection{Eligible Sets and Applicability}
\label{sec:eligible_set}

Our experiment measures drops in solve rate when content is removed from a report. For this to be meaningful, each instance in the working set must be solvable by the model under our experimental conditions. An instance the model cannot solve even with the full problem statement provides no information about the effect of removing content from it. We therefore construct a model-specific \textit{eligible set}. For each model, we run the full, unmodified problem statement on all 283 bug-fix instances three times and retain the instances resolved in at least one of the three runs. Since we select these instances precisely because the model solves them with the full problem statement, this set has a 100\% solve@3 rate at baseline, and every drop we report is measured against that baseline. Throughout, we count an instance as solved when at least one of its three runs passes, so each instance is judged by whether the model can still solve it rather than by the outcome of any single run. All spec-composition, localization, and structural mutations for a model are evaluated on that model's eligible set. Content mutations are measured against a separate report-content-only baseline. Because a content mutation removes a section from the report content, it is only informative on instances the model can already solve from report content alone. For each model, we build this second eligible set the same way, from the instances resolved in at least one of three report-content-only runs, and it likewise has a 100\% solve@3 rate at baseline. Not every component is present in every instance. The interface is optional, and report-content sections and file references do not appear in all instances, so each mutation is evaluated only on the instances that contain the information it targets. For the main conditions, we retain only eligible instances containing every component, giving the conditions a shared denominator. We report the applicable number of instances, $n$, for every condition, and compare each condition with its corresponding baseline on the same subset, which has a 100\% solve@3 rate by construction.

\subsection{Metrics}
\label{sec:study2_metrics}

We report two solve rates for every condition. The \textit{at-least-once} solve rate, reported as solve@3 in the tables, counts an instance as solved if at least one of its three runs resolved it. The \textit{mean solve} rate is the fraction of the three runs that passed, averaged over applicable instances. The mean solve rate captures a different signal, namely how \emph{reliably} a model solves an instance rather than merely whether it ever does. We express every condition's outcome as a drop in percentage points ($\Delta$\,pp) from its baseline, which is 100\% at-least-once by construction (Section~\ref{sec:eligible_set}). Because every condition starts from a 100\% baseline, each drop measures the same quantity: the share of solvable instances a mutation renders unsolvable. Conditions that share a denominator are therefore directly comparable, such as the spec-composition and structural conditions.
Conditions evaluated on different subsets, such as the interface, localization, and content mutations, are not directly comparable with those on the full eligible set. We interpret each drop on its own subset and do not rank its magnitude against a condition measured on a different subset.

\begin{table*}[t]
  \centering
  \caption{Study~2 ablation results.}
  \label{tab:study2-results}
  \footnotesize
  \setlength{\tabcolsep}{3pt}
  \renewcommand{\arraystretch}{1.2}
  \begin{tabular*}{\textwidth}{@{\extracolsep{\fill}}>{\raggedright\arraybackslash}p{1.7cm} l rrrr rrrr@{}}
    \toprule
    & & \multicolumn{4}{c}{Qwen} & \multicolumn{4}{c}{Gemma} \\
    \cmidrule(lr){3-6}\cmidrule(lr){7-10}
    \textbf{Family} & \textbf{Mutation} & $n$ & solved & solve@3 \% ($\Delta$) & mean solve \% ($\Delta$) & $n$ & solved & solve@3 \% ($\Delta$) & mean solve \% ($\Delta$) \\
    \midrule
    \rowcolor{gray!15} \multicolumn{2}{l}{\textit{Baseline: full statement}} & 79 & 79 & 100.0 & 53.6 & 83 & 83 & 100.0 & 70.7 \\
    \midrule
    \multirow{10}{*}{\shortstack[l]{\textbf{Spec-}\\\textbf{composition}}}
      & All $-$ Title          & 79 & 58 & 73.4 \textcolor{blue}{($-$26.6)} & 44.3 \textcolor{blue}{(\phantom{0}$-$9.3)}  & 83 & 75 & 90.4 \textcolor{blue}{(\phantom{0}$-$9.6)}  & 65.5 \textcolor{blue}{(\phantom{0}$-$5.2)}  \\
      & All $-$ Description    & 79 & 58 & 73.4 \textcolor{blue}{($-$26.6)} & 44.7 \textcolor{blue}{(\phantom{0}$-$8.9)}  & 83 & 71 & 85.5 \textcolor{blue}{($-$14.5)} & 63.5 \textcolor{blue}{(\phantom{0}$-$7.2)}  \\
      & All $-$ Report Content & 79 & 58 & 73.4 \textcolor{blue}{($-$26.6)} & 46.0 \textcolor{blue}{(\phantom{0}$-$7.6)}  & 83 & 67 & 80.7 \textcolor{blue}{($-$19.3)} & 55.4 \textcolor{blue}{($-$15.3)} \\
      & All $-$ Requirements   & 79 & 28 & 35.4 \textcolor{blue}{($-$64.6)} & 20.3 \textcolor{blue}{($-$33.3)} & 83 & 37 & 44.6 \textcolor{blue}{($-$55.4)} & 30.5 \textcolor{blue}{($-$40.2)} \\
      & All $-$ Interface      & 35 & 21 & 60.0 \textcolor{blue}{($-$40.0)} & 35.2 \textcolor{blue}{($-$18.4)} & 36 & 24 & 66.7 \textcolor{blue}{($-$33.3)} & 51.9 \textcolor{blue}{($-$18.8)} \\
      \cmidrule(lr){2-10}
      & Only Title          & 79 &  9 & 11.4 \textcolor{blue}{($-$88.6)} &  5.9 \textcolor{blue}{($-$47.7)} & 83 & 10 & 12.0 \textcolor{blue}{($-$88.0)} &  8.8 \textcolor{blue}{($-$61.9)} \\
      & Only Description    & 79 & 18 & 22.8 \textcolor{blue}{($-$77.2)} & 12.2 \textcolor{blue}{($-$41.4)} & 83 & 17 & 20.5 \textcolor{blue}{($-$79.5)} & 14.1 \textcolor{blue}{($-$56.6)} \\
      & Only Report Content & 79 & 20 & 25.3 \textcolor{blue}{($-$74.7)} & 14.3 \textcolor{blue}{($-$39.3)} & 83 & 27 & 32.5 \textcolor{blue}{($-$67.5)} & 20.9 \textcolor{blue}{($-$49.8)} \\
      & Only Requirements   & 79 & 54 & 68.4 \textcolor{blue}{($-$31.6)} & 39.7 \textcolor{blue}{($-$13.9)} & 83 & 63 & 75.9 \textcolor{blue}{($-$24.1)} & 54.6 \textcolor{blue}{($-$16.1)} \\
      & Only Interface      & 35 & 13 & 37.1 \textcolor{blue}{($-$62.9)} & 27.6 \textcolor{blue}{($-$26.0)} & 36 & 14 & 38.9 \textcolor{blue}{($-$61.1)} & 32.4 \textcolor{blue}{($-$38.3)} \\
     \midrule
    \textbf{Localization} & Remove All File References & 43 & 26 & 60.5 \textcolor{blue}{($-$39.5)} & 32.6 \textcolor{blue}{($-$21.0)} & 42 & 30 & 71.4 \textcolor{blue}{($-$28.6)} & 47.6 \textcolor{blue}{($-$23.1)} \\
    \midrule
    \multirow{2}{*}{\textbf{Structural}}
      & Remove Section Headers & 79 & 55 & 69.6 \textcolor{blue}{($-$30.4)} & 47.7 \textcolor{blue}{(\phantom{0}$-$5.9)} & 83 & 75 & 90.4 \textcolor{blue}{(\phantom{0}$-$9.6)}  & 67.1 \textcolor{blue}{(\phantom{0}$-$3.6)} \\
      & Flatten Lists          & 79 & 58 & 73.4 \textcolor{blue}{($-$26.6)} & 45.1 \textcolor{blue}{(\phantom{0}$-$8.5)} & 83 & 74 & 89.2 \textcolor{blue}{($-$10.8)} & 64.7 \textcolor{blue}{(\phantom{0}$-$6.0)} \\
    \midrule
    \rowcolor{gray!15} \multicolumn{2}{l}{\textit{Baseline: report content only}} & 20 & 20 & 100.0 & 56.7 & 26 & 26 & 100.0 & 66.7 \\
    \multirow{4}{*}{\shortstack[l]{\textbf{Report}\\\textbf{Content}}}
      & RC $-$ Observed Behavior  & 20 & 12 & 60.0 \textcolor{blue}{($-$40.0)} & 30.0 \textcolor{blue}{($-$26.7)} & 26 & 21 & 80.8 \textcolor{blue}{($-$19.2)} & 52.6 \textcolor{blue}{($-$14.1)} \\
      & RC $-$ Expected Behavior  & 20 & 13 & 65.0 \textcolor{blue}{($-$35.0)} & 43.3 \textcolor{blue}{($-$13.4)} & 26 & 15 & 57.7 \textcolor{blue}{($-$42.3)} & 46.2 \textcolor{blue}{($-$20.5)} \\
      & RC $-$ Steps to Reproduce & 13 &  8 & 61.5 \textcolor{blue}{($-$38.5)} & 38.5 \textcolor{blue}{($-$18.2)} & 12 &  9 & 75.0 \textcolor{blue}{($-$25.0)} & 66.7 \textcolor{blue}{(\phantom{$-$0}0.0)}    \\
      & RC $-$ Additional Context &  8 &  5 & 62.5 \textcolor{blue}{($-$37.5)} & 50.0 \textcolor{blue}{(\phantom{0}$-$6.7)}  & 12 &  9 & 75.0 \textcolor{blue}{($-$25.0)} & 58.3 \textcolor{blue}{(\phantom{0}$-$8.4)}  \\
    \bottomrule
  \end{tabular*}
\end{table*}

\subsection{Results}
\noindent\textbf{RQ2: How do controlled changes to the content and structure of a problem statement affect agent solve rates?}

Table~\ref{tab:study2-results} reports the ablation results. Spec-composition either removes one component from the full statement (All\,$-$\,\textit{X}) or keeps only that component (Only~\textit{X}); Localization removes file references; and Structural reformats the statement by flattening lists or removing section headers. Each condition measures how many instances stay solved. Report Content removes one part of the report content and compares to a separate report-content-only baseline. We report solve@3 and mean solve, each with its change ($\Delta$) from the baseline.

\textbf{Spec-composition mutations.} The three natural parts of the bug report, the title, description, and report content, were equally important in aggregate. For Qwen, removing any one of them left the same solve@3 rate of 73.4\% (58 of 79), a 26.6 percentage-point drop, and their mean solve values were close as well, 44.3\% without the title, 44.7\% without the description, and 46.0\% without report content, against 53.6\% for the full statement. They solved the same number of instances but not the same instances, with a pairwise Jaccard of 0.26 to 0.34 on the unsolved bugs, so each deletion lost a partly different set of bugs. Gemma showed the same general result with one difference. Removing report content (80.7\%) hurt slightly more than removing the title (90.4\%) or the description (85.5\%), and the mean solve rates followed the same order, so report content mattered somewhat more for Gemma.

None of the three parts was sufficient on its own. Alone, the title solved 11.4\% of instances for Qwen and 12.0\% for Gemma, the description 22.8\% and 20.5\%, and report content 25.3\% and 32.5\%, with mean solve rates lower still, ranging from 5.9\% to 20.9\%. These low mean rates put the solve@3 numbers in context. When the full statement is cut down to an incomplete version, repair becomes highly unstable. The task does not become impossible to solve; it makes the missing parts something the agent has to infer, and across runs the model forms different hypotheses about the intended behavior or where to look. A passing patch under solve@3 can therefore mean that one run happened to reconstruct enough of the missing specification while the others did not. 

The two parts that SWE-bench Pro adds on top of the report, the requirements and the interface, produced larger drops than any other component, but neither is something a reporter writes. The benchmark derives both from the resolved task, so they state the behavior a correct patch must satisfy and the files it must touch. Removing the requirements cut solve@3 by 64.6 percentage points for Qwen and 55.4 for Gemma, and requirements alone solved more instances than any other single component, 68.4\% for Qwen and 75.9\% for Gemma. Removing the interface cut solve@3 to 60.0\% for Qwen and 66.7\% for Gemma on the subsets that carried one. We separate these from the question of what a good report contains, since they are not part of a natural bug report, and report them only to show how much both models gain from a precise specification when one is supplied.

\textbf{Localization information.} Removing every reference to gold-patch files lowered Qwen's solve@3 to 60.5\% (26 of 43), a 39.5 percentage-point drop. Gemma retained 30 of 42 instances (71.4\%), a 28.6-point drop. In most broken runs across both models, the agent wrote the correct code but put it in the wrong file, the same wrong-file error the SWE-bench Pro authors report as a common failure mode \cite{deng2025swe}. When the location is removed, the agent has to infer where the change belongs. It can sometimes recover on its own if the rest of the report still points to the right code, but fails when the fix introduces a new symbol or could plausibly live in one of several similar modules.

\textbf{Structural mutations.} The organization of the report mattered. Flattening lists retained 73.4\% of Qwen instances and removing section headers retained 69.6\%, even though neither mutation removed task information. Gemma was less sensitive, retaining 89.2\% and 90.4\% respectively. That a drop occurs at all, when no task information was removed, shows the model depends on how the report is organized, not just its content. Trajectory inspection shows why. Flattening the numbered steps to reproduce and requirements into prose led Qwen to implement only part of the specification, and all three runs exhausted the 50-turn budget without a correct patch where the unmutated variant had solved it in two of three runs. Removing section headers similarly caused confusion between observed and expected behavior. Stripping the \emph{Observed Behavior} label blurred the symptom and the intended behavior, and in some cases the model submitted a patch based on the conflated interpretation in all three runs even after running tests.

\textbf{Report-content mutations.} The models differed more clearly inside report content.  We start with instances the models could solve from report content alone, then remove one section at a time. For Qwen, removing observed behavior reduced solve@3 from 100\% to 60.0\% and expected behavior to 65.0\%, while removing steps to reproduce retained 61.5\% and additional context 62.5\%. The applicable sets differ across these conditions, so the numbers do not support a ranking among the four sections, but every deletion caused a large drop of 35.0 to 40.0 percentage points from its matched report-content-only baseline. Once Qwen has to rely on report content alone, no single part of it is safely removable. Gemma instead relied slightly more on one section. Removing expected behavior caused its largest drop, from 100\% to 57.7\%, whereas removing observed behavior retained 80.8\%, partly because observed behavior was often restated in the surviving steps to reproduce. We examine the trajectories to understand what the drops mean and why the two models diverge. In the failed expected-behavior runs, Gemma commonly found the relevant code and made a plausible change, but the remaining report allowed more than one plausible interpretation of the task, and Gemma committed to one without searching further to resolve the ambiguity. Qwen, on the other hand, attempted to recover the missing section by searching more widely, writing its own reproduction script, or running additional tests, which lengthened its runs, and when observed behavior or steps to reproduce was removed, its submission rate fell as more runs reached the 50-turn limit without submitting a patch. In both cases the missing section changed how the model searched and the course of the entire repair loop, not only what it could solve.

\section{Discussion}
\label{sec:discussion}

\subsection{Is a good bug report for a human a good bug report for an agent?}

Table~\ref{tab:comparison} places our findings beside the features that human-centered studies identify as most valuable. The two overlap, but only in part. On some features, the agent and the human have contrasting needs. The clearest contrast is in the reproduction of the bug. A written list of the steps that trigger a bug was the most useful field for the human developers surveyed by Zimmermann et al.~\cite{zimmermann2010makes}. For an agent, the same written steps showed no association with success, while a reproduction script the agent can run was among the features that helped most. The two describe closely related information, both conveying how to trigger the bug, but in different forms: written steps for a person to follow and code for an agent to execute. Length and readability are also features where agents and humans differ. Zimmermann et al.~\cite{zimmermann2010makes} report that longer reports correlated with higher developer quality ratings, which they attribute to more information in the reports. Easier-to-read reports were resolved faster, with fixed reports exhibiting simpler readability scores than others. For repair agents, longer reports were associated with lower odds of resolution, and reports that scored as harder to read had slightly higher odds. A report that reads as more complex usually does so because it carries more technical detail, which seems to help agents. However, the extra material in longer reports could be noise that leads the agent down the incorrect repair path. More details about the similarities or differences in the bug report information needs between humans and agents are in Table~\ref{tab:comparison}.

\begin{table*}[t]
\centering
\footnotesize
\setlength{\tabcolsep}{4pt}
\renewcommand{\arraystretch}{1.25}
\caption{Comparison of bug report features across our two studies (RQ1, RQ2) and prior human studies.}
\label{tab:comparison}
\begin{tabularx}{\textwidth}{@{}>{\raggedright\arraybackslash}p{2.4cm} >{\raggedright\arraybackslash}p{4cm} >{\raggedright\arraybackslash}p{5.5cm} >{\raggedright\arraybackslash}X@{}}
\toprule
\textbf{Feature} &
\textbf{RQ1 — Statistical Associations} \newline
{\normalfont\itshape SWE-bench Verified, 87 agents} &
\textbf{RQ2 — Causal Effects via Ablation} \newline
{\normalfont\itshape SWE-bench Pro, 2 models, solve@3 $\Delta$pp} &
\textbf{Past Studies} \newline
{\normalfont\itshape Developer-centric findings} \\
\midrule
\textbf{Observed behavior} & Excluded; near-universal (98.8\%). & Removal causes a 19--40\,pp drop. Qwen is more sensitive ($-$40\,pp). Gemma less so ($-$19\,pp), as OB may often be inferable from other sections. & Surveys indicate 33\% of developers rank OB as most helpful and it is often provided\cite{bettenburg2008makes}. Considered a critical section excluding which wastes developer time \cite{bo2024chatbr}\\
\textbf{Expected behavior} & Excluded; near-universal (91.9\%). & Removal causes largest drop for Gemma ($-$42\,pp) compared to Qwen ($-$35\,pp). Without EB, models commit to ambiguous interpretations and exhaust their turn budget. & Another critical section (used by 89\%) and ranked most helpful by 22\% \cite{bettenburg2008makes}. Despite its perceived usefulness, data shows only 35.2\% of bug reports explicitly include EB \cite{chaparro2017detecting} \\
\textbf{Steps to reproduce} & No association (OR 0.83). & Removal causes a 25--39\,pp drop. Agents spend more turns trying to reproduce bugs, increasing failure by exhausting turn budget. & Rated most useful and helpful by developers \cite{bettenburg2008makes, davies2014s}. Lack causes excessive manual effort during triage and result in non-reproducible and unfixed bugs \cite{chaparro2019assessing}. \\
\textbf{Stack traces / error messages} & No association (OR 0.93/1.12). & Not ablated as a standalone section. & Considered important for software debugging \cite{soltani2020significance}. Up to 60\% of fixed bugs involved changes directly within the stack frames mentioned, and they shorten bug resolution times \cite{schroter2010stack, zimmermann2010makes} \\
\textbf{Code snippets} & No association (OR 1.45). & Not ablated as a standalone section. & Perceived as having low importance but harder to provide \cite{soltani2020significance}. Reportedly increases chances of the report getting fixed \cite{zimmermann2010makes}. \\
\textbf{Repository code} & Positive (OR 2.82*). & Not ablated as a standalone section. &  Presence weakly correlates with a bug being fixed. May provide little significant benefit to look into investigated aspects \cite{davies2014s}.\\
\textbf{Reproduction scripts} & Positive (OR 2.52**). & Not tested. Although agents spent turns creating their own reproduction scripts. &  Considered to be more in line with developer needs, unlike prose which can be vague and lacking key details \cite{rahman2020some}\\
\textbf{Fix suggestions} & Strongest positive (OR 3.61***). & Fix guidance in Requirements (55--65,pp) and Interface (33--40,pp) removal had a large effect; it also sometimes appeared in Additional Context (25--38,pp). & Considered to be low importance but inclusion shows statistically significant improvement in bug resolution time \cite{soltani2020significance}.\\
\textbf{Readability} & Harder prose (higher SMOG), higher success (OR 1.55**). & Not tested. & Readability improves perceived quality and shorter bug resolution times \cite{bettenburg2008makes, hooimeijer2007modeling} \\
\textbf{Lists / headers} & No association (OR 0.82/1.78). & Removal causes 10--30\,pp drops.  Without headers, agents confuse OB and EB. Implementing only part of the specification. & Itemization is looked at as a quality signal \cite{zimmermann2010makes}. It adds structure and helps developers read and understand better \cite{bettenburg2008extracting} \\
\textbf{Report length} & Longer, lower success (OR 0.49***). & Not isolated from content removal. & Longer descriptions tend to have more information leading to faster resolution \cite{dal2016makes}. \\
\textbf{Fault localization} & Positive (OR 2.33*). & Removing file references causes a 29--40\,pp drop. Agent ends up patching the wrong file. &  Locations to fix perceived useful \cite{zimmermann2010makes} but harder to provide \cite{dal2016makes}.\\
\bottomrule
\end{tabularx}
\end{table*}

\subsection{So what makes a good bug report for an agent?}

Both study 1~(Section~\ref{sec:study1_statistical_modeling}) and study 2 (Section~\ref{sec:study2_ablation}) point to the same answer: \emph{A report serves an agent well when it reduces what the agent must work out on its own}. The reason lies in how an agent differs from a human reader. A developer who finds a report incomplete
can ask the reporter or keep investigating before committing to a change. An agent currently does not. When the report is missing something, the agent does not stop to ask. It infers what is missing and acts on that inference, whether or not the inference is correct. 
We discuss some implications for reporters on what is important in a bug report for an agent:

\textbf{\textit{Point the agent to the fix location.}} Fault localization helped in both studies. A report that named a file changed by the fix was associated with success in Study 1, and removing those references in Study 2 lowered the solve rate for both models. 
The reason could be based on how developers and agents differ when it comes to fixing bugs. If a developer does not know where to fix a bug, they will not go ahead and add a fix in a random location. They would rather not fix the bug than fix it in an incorrect location. An agent, however, cannot say that it can't fix a bug. When tasked with a repair task, it will always produce a fix, even if it does not know where to include the fix. 
In Study 2, the most common mistake was a correct change ending up in the wrong file. Therefore, \emph{any clue that narrows the search, such as a file path, a function name, or a unique error string, prevents the agent from applying the correct fix in an incorrect location}.

\textbf{\textit{Give the agent an executable reproduction script.}} Both studies showed that a reproduction script has a larger impact on an agent compared to the reproduction steps in natural language. 
While both may carry similar information, the agent can run a script directly. We find that the agent often made mistakes when it tried to create such an executable script from the described steps. Therefore \emph{providing a reproduction script that the agent can run removes mistakes that can happen when trying to recreate an executable script from the natural language steps}.

\textbf{\textit{Keep the report structured.}} The organization of the report mattered as well, independent of its content. Removing section headers and flattening lists lowered solve rates in Study 2 even though neither edit changed any text, with a stronger effect for Qwen than
for Gemma. Section headers keep the parts of a report apart. When the Observed Behavior label was removed, the agent confused the reported symptom with the behavior the program should produce, and in some runs patched the confused
reading even after running the tests. Flattening a numbered list had a similar effect, and the agent implemented only part of the steps or requirements that were described. Because these edits removed no information, the drops indicate that \emph{the agent relies on how the information in a report is organized and not only on what information is stated}.

\textbf{\textit{Models fill gaps differently.}} The ablation showed how removing information changes how the agent works. 
When a section was missing, the agent did not simply fail on the instances. It changed how the repair was approached. That change ran through the rest of the repair. 
Our analysis shows how the choice of the model could impact the agent attempting the repair with partial information. Qwen treated a missing section as something to recover. It searched the repository more widely, wrote its own reproduction, and ran more tests, which lengthened its attempt and, under some conditions, used up the 50-turn budget before it submitted a patch. Gemma tended to commit to a change early, often without running it, and when the remaining report supported more than one reading of the task, Gemma chose one and stopped. When Gemma failed, it was often a confident patch built on the wrong reading of the report. The same deletion, therefore, had a different effect depending on the model. Removing the expected behavior left Gemma free to act on a plausible but unintended guess, while Qwen went into a
longer search to pin the behavior down. 
Therefore, \emph{the report has to be explicit enough that a model inclined to guess is not left to act on the guess}, even though some models~(like Qwen) may not.

\section{Threats to Validity}
\label{sec:threats_validity}

\emph{Internal Validity:} Study~1 is observational and cannot establish that any feature directly causes resolution success or failure. We control for bug difficulty and differences among agents, but unmeasured factors may remain. An LLM generated initial labels, which two authors independently reviewed and verified; Cohen's $\kappa$ ranged from 0.92 to 1.00, though residual labeling errors remain possible. We address non-determinism by repeating each condition three times and using solve@3. Study~1 uses SWE-bench Verified, which predates some models' training cutoffs, so leaked fixes could let an AI repair agent solve without relying on the report and affect apparent feature effects. SWE-bench Pro mitigates this concern, and the two studies also test whether the same patterns hold across benchmarks.

\emph{Construct Validity:} We define resolution by benchmark test execution, following the standard SWE-bench protocol. Passing tests do not guarantee a fully correct fix, and may incorrectly reject valid solutions, a limitation shared by all such benchmarks. Feature presence is annotated as a binary value, hence does not capture the quality of the underlying data. A one-line fix suggestion and a detailed root-cause analysis are treated identically under this scheme. 

\emph{External Validity:} Both studies draw on open-source GitHub issues and may not generalize to industrial reports. Study~2 evaluates two open-weight model families with a minimal scaffold, so results may differ for proprietary models or other agent harnesses. Full-scale proprietary evaluation was infeasible, but on 10 instances \texttt{Claude Sonnet 4.6}  \cite{anthropic2026sonnet46} showed the same pattern: no component alone sufficed, title-only and description-only each solved 3 of 10 instances, and removing observed or expected behavior also left 3 of 10 solved. This supports the finding that AI repair agents degrade sharply when core report content is missing or incomplete.

\section{Conclusion}
\label{sec:conclusion}

Agents and humans have differing information needs from bug reports. Any information that would lead an agent to guess will cause it to make mistakes, therefore the clearer a bug report is, the better it is for an agent. The reports that helped most named the file the fix changes, included a script that reproduces the failure, and suggested a direction to take. This distinction matters as repair agents move from benchmarks into everyday development. Bug-tracking systems, report templates, and reporting guidelines were all built around what a human developer needs, and much of that effort went toward prose that reads clearly and steps a person can follow. 

Until agents can reason about requirements, collect missing details and make informed decisions the way a developer does, the burden falls on the report to provide the information agents need to reliably fix the bugs reported.

\section{Data Availability}

Our datasets, scripts, and full experimental results are available at the following online repository: \url{https://github.com/uw-swag/agent-bug-reports}

\bibliographystyle{IEEEtran}
\bibliography{references}

\end{document}